\documentclass[12pt]{article}

\usepackage{amsmath}
\usepackage{amssymb}
\usepackage{amsfonts}
\usepackage{graphicx}

\numberwithin{equation}{section}

\begin{document}

\begin{center}
{\large \bf{ The Higgs condensate as a quantum liquid: Comparison with the ATLAS full Run 2  data }}
\end{center}

\vspace*{2cm}

\begin{center}
{
Paolo Cea~\protect\footnote{Electronic address:
{\tt paolo.cea@ba.infn.it}}  \\[0.5cm]
{\em INFN - Sezione di Bari, Via Amendola 173 - 70126 Bari,
Italy} }
\end{center}

\vspace*{1.5cm}

\begin{abstract}
\noindent 
Recently, we proposed to picture the Higgs condensate of the Standard Model as a quantum liquid analogous to
the superfluid Helium II.  In this scenario  the Higgs condensate excitations resemble closely two Standard Model Higgs bosons.
The lightest  Higgs boson was  already identified with the LHC narrow resonance at 125 GeV. Concerning 
the heavy Higgs boson, we  found preliminary evidence in our previous phenomenological analysis in the so-called golden
channel. In the present note we compared our proposal to the full Run 2 data set released  recently by the ATLAS Collaboration.
Even though  we do not found a clear statistical evidence for our Standard Model heavy Higgs,  we found  that
our theoretical proposal is still in accordance with the available  observations.
\vspace{0.5cm}

\vspace{0.2cm}


\end{abstract}
\newpage
\noindent
\section{Introduction}
\label{s-1}
In a recent article~\cite{Cea:2020} we advanced the proposal that  the Higgs condensate of the Standard Model  should be considered
like a  relativistic quantum fluid analogous  to superfluid helium. As discussed at length in Ref.~\cite{Cea:2020}, we found that
 there are two different kind of Higgs condensate excitations that are similar to phonon and rotons in helium II. Moreover,
 in the dilute gas approximation these two Higgs condensate excitations behave like the Standard Model  Higgs 
 boson~\cite{Englert:1964,Higgs:1964,Guralnik:1964,Higgs:1966} and an heavy Higgs boson with mass around 730 GeV. 
Interestingly enough, it is noteworthy that an alternative scenario for the two-Higgs picture has been recently
advanced  in Refs.~\cite{Consoli:2020a,Consoli:2020b}. \\
Previous analyses of the preliminary LHC Run 2 data in the so-called golden channel  $H \rightarrow \ell^+ \ell^- \ell'^+ \ell'^-$ where $\ell , \ell' = e$ or $\mu$,  seems to favour  some evidence of a broad scalar resonance with mass around 700 GeV that looks  consistent with a heavy Higgs boson~\cite{Cea:2017,Cea:2019,Richard:2020}.
The aim of the  present note is to upgrade our previous analyses to the full LHC Run 2 data set  from the ATLAS Collaboration,
considering that the CMS Collaboration has not yet released  the Run 2 data at least in the phase-space regions relevant to
our purposes.  In particular we shall contrast our  theoretical proposal  to the latest ATLAS data  assuming that the additional
heavy Higgs boson (denoted as H) is produced via gluon-gluon fusion  (GGF) processes in Sect.~\ref{s-2},  or vector-boson fusion (VBF) processes
in Sect.~\ref{s-3}. Finally, in Sect.~\ref{s-4} we summarise the main results of the paper.
\section{The gluon-fusion production mechanism}
\label{s-2}
\begin{figure}
\vspace{-0.0cm}
\includegraphics[width=0.5\textwidth,clip]{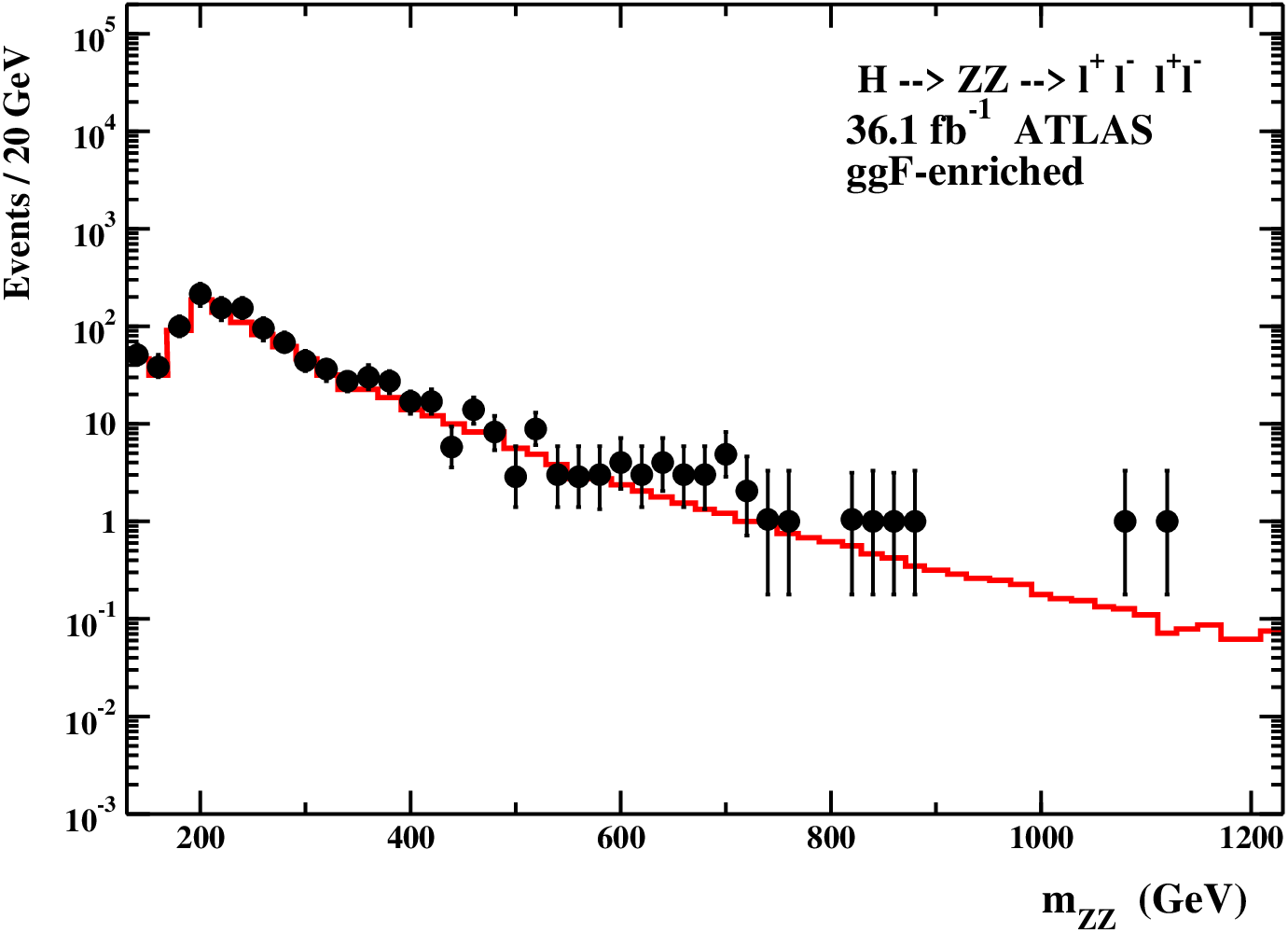}
\hspace{0.2 cm}
\includegraphics[width=0.5\textwidth,clip]{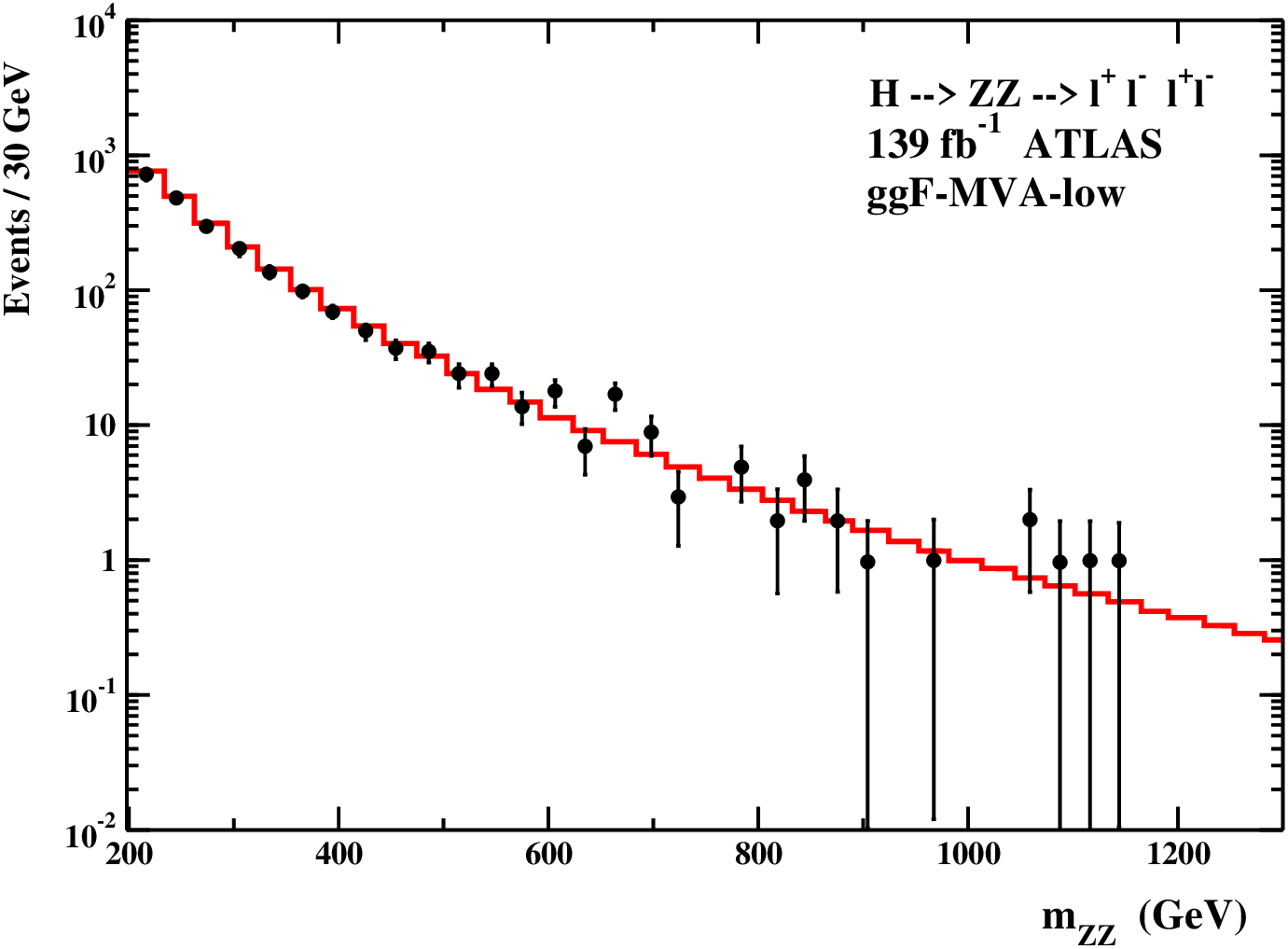}
\caption{\label{Fig1} 
Distribution of the invariant mass $m_{Z Z}$ for the GGF  processes $H \; \rightarrow ZZ \; \rightarrow \ell \ell \ell \ell$   ($\ell = e, \mu$)
corresponding  to  an  integrated luminosity of  ${\cal{L}} = 36.1 \,  fb^{-1}$ (left panel) and $139 \,  fb^{-1}$ (right panel).
The data  have been obtained  from Fig.~4 (a) in Ref.~\cite{Aaboud:2018} and  Fig.~2  (d) in Ref.~\cite{Aad:2021},  respectively.  
The (red) continuous lines  are the ZZ irreducible backgrounds.
}
\end{figure}

In our previous papers~\cite{Cea:2020,Cea:2019} we showed that for large Higgs masses the main
production processes are by vector-boson fusion and gluon-gluon fusion processes. Moreover,
 almost all the  decay modes of the  heavy Higgs boson H are given by the decays into W$^+$W$^-$ and Z$^0$Z$^0$ with:
\begin{equation}
\label{2.1}
Br(H \rightarrow W^+ W^-) \; \simeq \; 2 \; Br(H \rightarrow Z^0 Z^0) \; .
\end{equation}
It is well established that  the decay channels $H  \rightarrow  ZZ  \rightarrow  4 \ell$ (the golden channel) have very low branching ratios,
nevertheless  the presence of leptons allows to efficiently reduce the huge background due mainly to diboson production.
Indeed, the four-lepton channel, albeit rare, has the clearest and cleanest signature of all the
possible Higgs boson decay modes due to the small background contamination. 
Usually, it is assumed that an additional Higgs boson would be produced
predominantly via gluon-gluon fusion (GGF) and vector-boson fusion (VBF) processes. Therefore,  the
events are classified into GGF and VBF categories and results are interpreted separately for the GGF and VBF production modes.
A search for a new high-mass resonances decaying into electron or muon pairs has been performed 
the ATLAS  experiment  using data collected at $\sqrt{s} = 13$  TeV  corresponding to an integrated luminosity of 
36 fb$^{-1}$~\cite{Aaboud:2018} and upgraded to 139 fb$^{-1}$~\cite{Aad:2021}. \\
In Fig.~\ref{Fig1}  we show the invariant mass distribution for the golden channel for the GGF category corresponding to
36 fb$^{-1}$ (left panel) and 139 fb$^{-1}$ (right panel).
In order to be sensitive to the VBF production mode, for the 36.1 $fb^{-1}$ data set, 
the ATLAS Collaboration~\cite{Aaboud:2018} classified the events into four categories, namely  one for the
VBF production mode and three for the GGF production mode.
 If an event has two or more jets with $p_T$ greater than 30 GeV, with the two leading jets $j_1,j_2$ well separated in
the pseudo-rapidity $\eta$,  i.e. $| \eta_{j_1j_2} |  > 3.3 $,  and having an invariant mass $m_{j_1j_1}  > 400$  GeV, 
then this event is classified into the VBF-enriched category.
 Otherwise the event is classified into one of the GGF-enriched categories. Such classification is
used only in the search for a heavy scalar produced with the narrow width approximation. \\
On the other hand, for the full Run 2 data set the event classification targeting different production processes has been optimised
using machine learning algorithms~\cite{Aad:2021}.
 More precisely, to improve the sensitivity in the search of a heavy Higgs signal produced either in the VBF 
or in the GGF production mode, were used two classifiers,  a  VBF classifier  and a GGF classifier. 
These classifiers were built with deep neural networks. The networks were trained  by means of  simulated signal events
from a heavy Higgs boson with masses ranging from 200 GeV up to 1400 GeV in the narrow width approximation,
 and from the Standard Model continuous  ZZ background. \\
\begin{figure}
\vspace{-0.0cm}
\includegraphics[width=0.5\textwidth,clip]{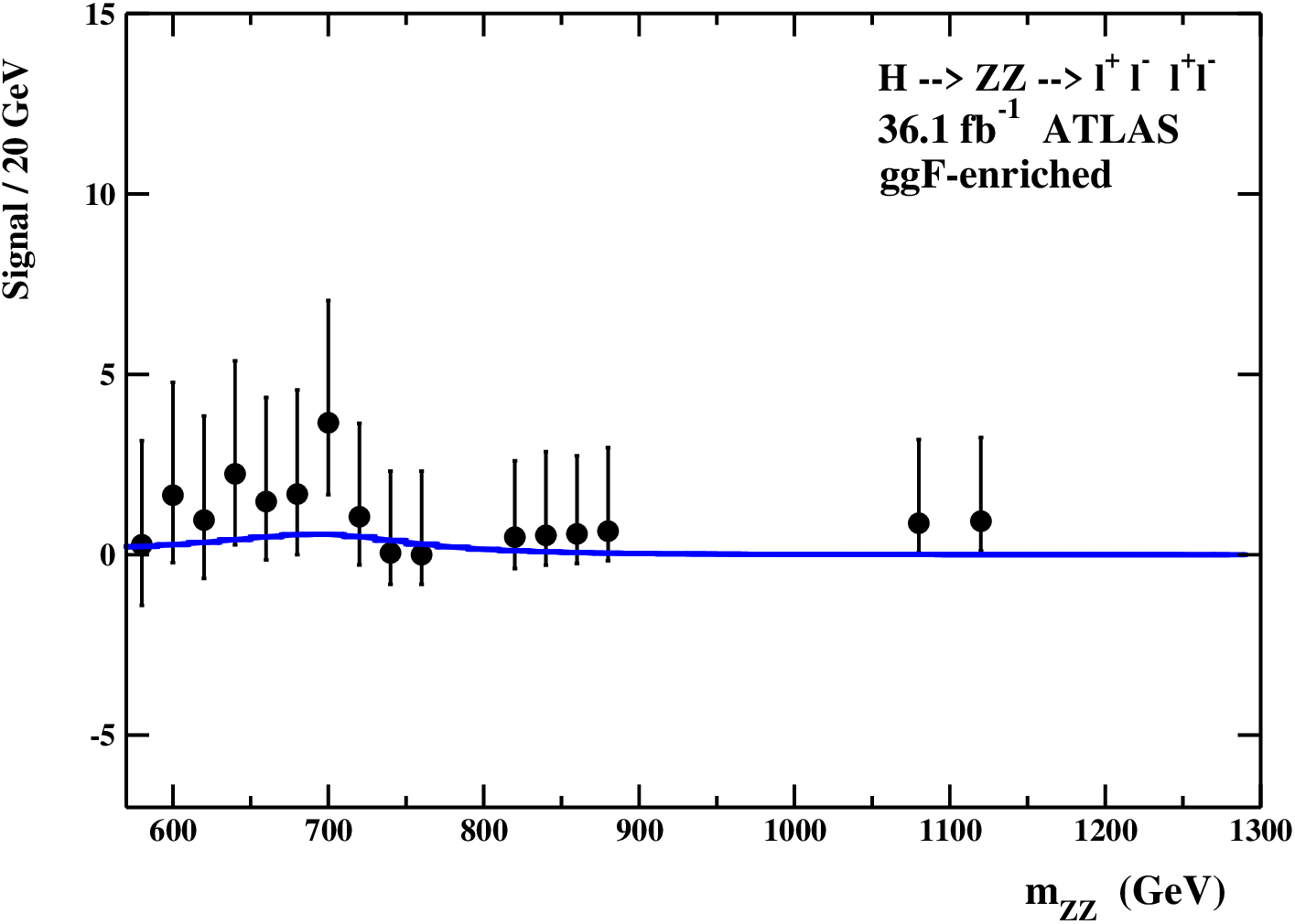}
\hspace{0.2 cm}
\includegraphics[width=0.5\textwidth,clip]{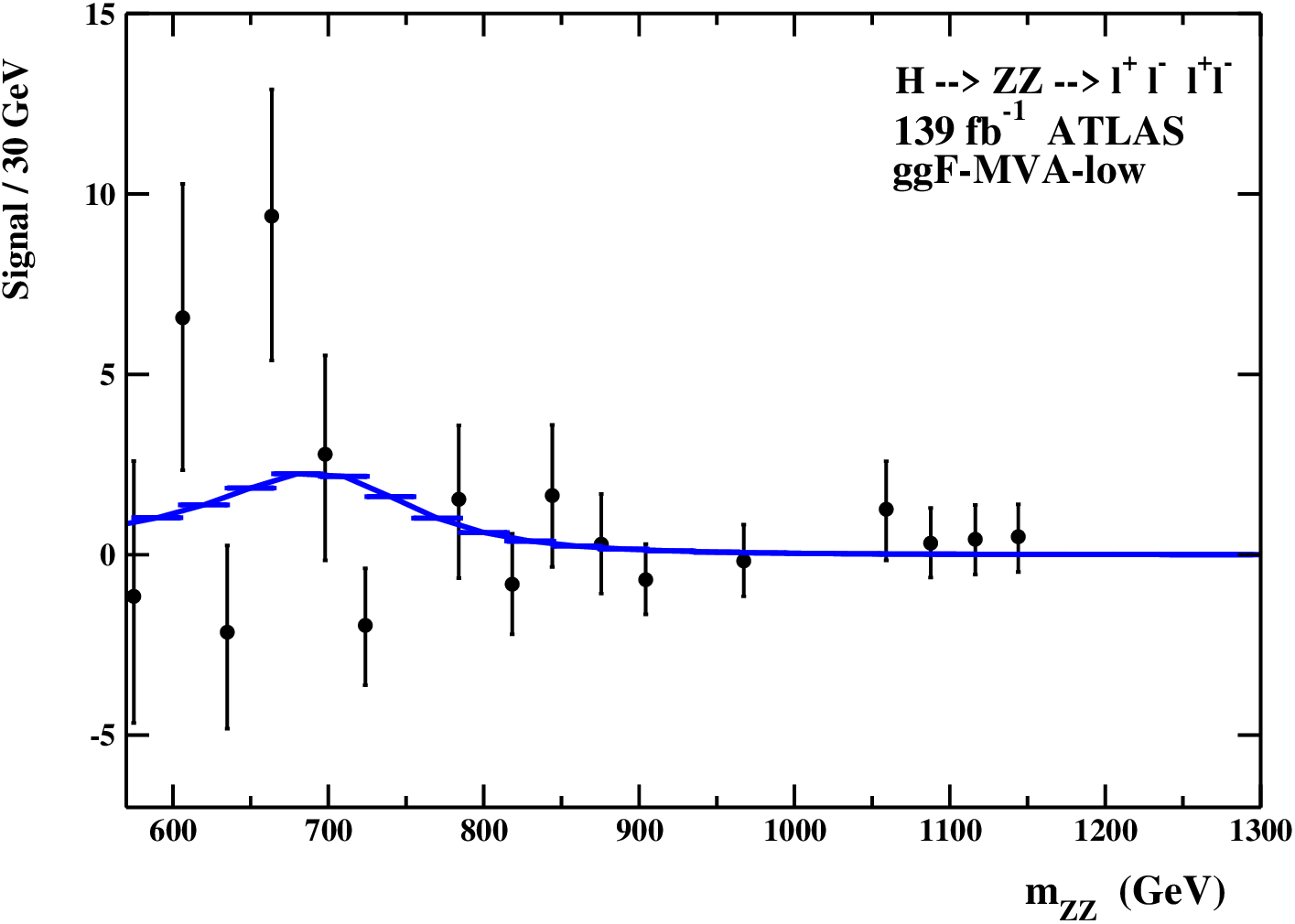}
\caption{\label{Fig2} 
Comparison to the Run 2  data of the distribution of the  invariant-mass $m_{Z Z}$  distribution  
in the high-mass region   $m_{Z Z}  \gtrsim 600 \,  GeV$  for the  GGF processes
 $H \; \rightarrow ZZ \; \rightarrow \ell \ell \ell \ell$   ($\ell = e, \mu$) with integrated luminosity 36.1 fb$^{-1}$ (left panel)
 and 139 fb$^{-1}$ (right panel). The signal distribution  has been obtained from the relevant  ATLAS event distributions  by
subtracting the ZZ backgrounds.  
 }
\end{figure}
From the event distributions we may easily obtain the signal distributions by subtracting the continuous  irreducible  ZZ backgrounds
that constitute the main source of the Standard Model background in the invariant high-mass region. The results are displayed
in Fig.~\ref{Fig2}.  After that, we compare the observed signal distributions to our theoretical proposal. For the 2016 ATLAS data set,
we see that the expected signal histogram is perfectly compatible with the data, but it is evident from Fig.~\ref{Fig2}, left panel,
that the integrated luminosity is too low to claim an evidence of our heavy Higgs boson.
As concern the full data set, in Fig.~\ref{Fig2}, right panel, we compare the expected signal distribution  to the data.
For the comparison we have taken into account that the selection cuts applied by the ATLAS Collaboration to
the full Run 2 data set are more stringent with respect to the preliminary data. This can be seen by looking at the number of events
in the the high-mass region   $m_{Z Z}  > 600 \, $ GeV.  From the 36.1 fb$^{-1}$ data we infer that there  are about  28 events.
Since  the luminosity  increases by a factor $139/36.1 \simeq  3.85$, we should have about 108 events. Actually, we found that
the full Run 2 data set has about  75 events in the high-mass region. To take care of this we introduce an effective
efficiency factor:
\begin{equation}
\label{2.2}
\eta_{eff}^{GGF} \; \simeq \;  \frac{75}{108} \; \simeq \; 0.70 \; \; ,  
\end{equation}
and scaled the production cross section accordingly.  After that we compare the theoretical distribution to the data.
Looking at Fig.~\ref{Fig2}, right panel, we see that the data do display some broad structure around  $m_{Z Z}  \simeq 700 $ GeV
that compares reasonably with the theoretical expectations.
\begin{figure}
\vspace{-0.3cm}
\begin{center}
\includegraphics[width=0.66\textwidth,clip]{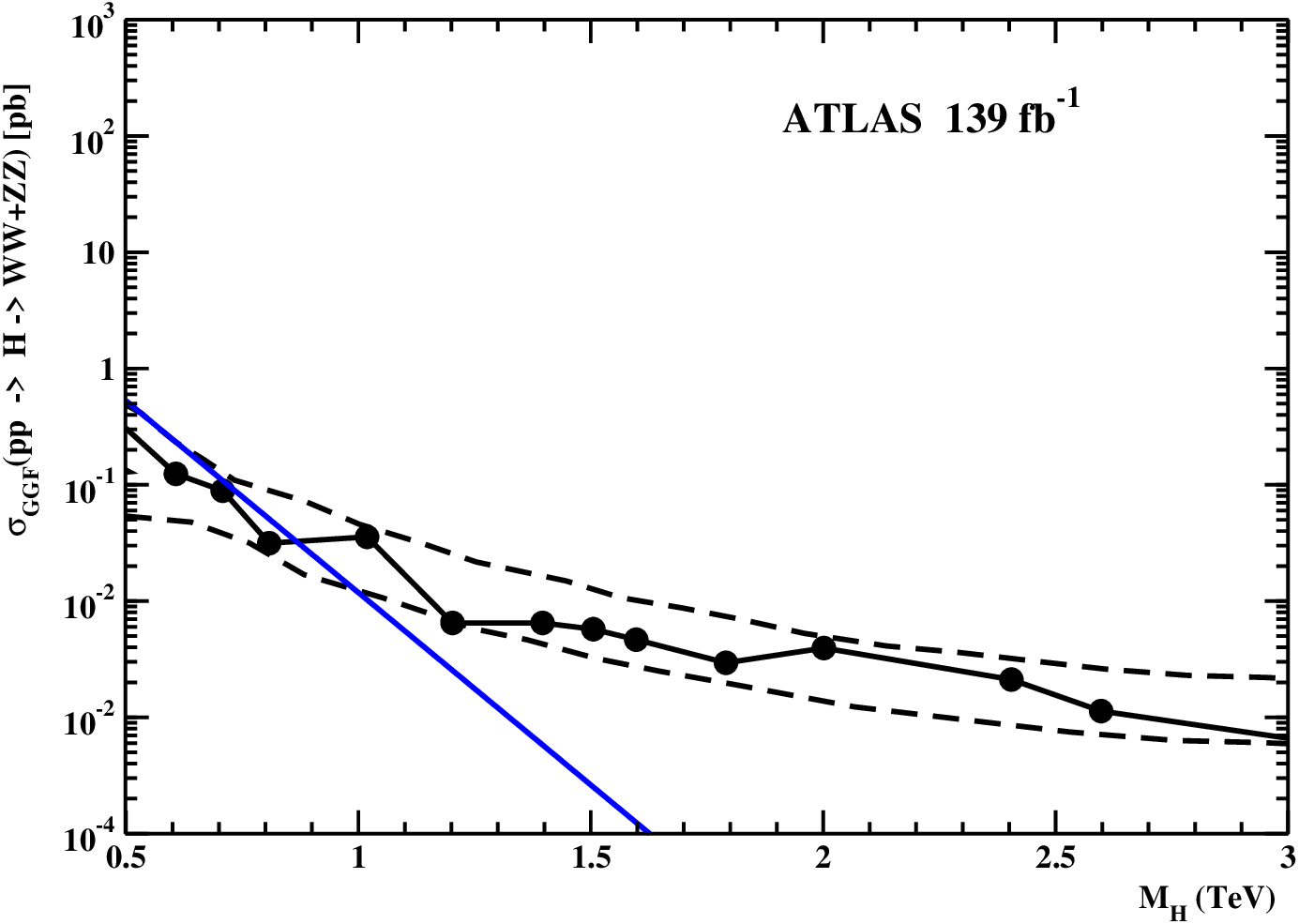}
\end{center}
\caption{\label{Fig3} 
 Limits on the GGF production cross section times the branching fraction for the processes $pp \rightarrow H \rightarrow VV$, V = W,Z. 
The data have been taken from Fig.~12 (a) in Ref.~\cite{Aad:2020}. The dashed  lines demarcate  the 95 \% confidence level region
of the expected Standard Model background. Full circles represent the observed signal. The continuous (blu) line
is our theoretical estimate for the gluon-gluon fusion production cross section times the branching ratio 
Br(H $\rightarrow$ WW  + ZZ).
}
\end{figure}
To be quantitative, we may estimate  the total number of events in the invariant mass interval  
 $ m_{ZZ}  >  700$  GeV and compare with our theoretical expectations.
We find:
\begin{equation}
\label{2.3}
N^{exp}_{sign} ( m_{ZZ}  > 600 \,  {\text GeV}) \; \simeq  \; 19.68^{+7.44 }_{- 8.52}      \; , \; 
N^{th}_{sign} =  12.20   \; \;  \;  \;  \; \;   \;  GGF  \;  \;  {\cal{L}} = 139 \,  {\text fb}^{-1}
\end{equation}
where the quoted errors have been obtained by adding in quadrature the 
experimental errors.\footnote{Strictly speaking, the combination of measurements with asymmetric errors should be
implemented by  combining the  likelihood functions (for instance, see Ref.~\cite{Barlow:2020}).  Here, we are adopting a conservative
attitude for  adding errors in quadrature  leads, in general, to an overestimate of the statistical errors.}
Even thought  the observed and predicted event counts are in quite good agreement, the background-only  hypothesis is
still consistent with observations, albeit at about two standard deviations. \\
According to Eq.~(\ref{2.1}),  the main decay mode of a heavy Higgs boson is the decays into two W vector bosons. Thus,
the most stringent constraints should arise from the experimental searches for a heavy Higgs boson
decaying into two W gauge bosons. In fact, in our previous paper~\cite{Cea:2020} we compared our theoretical expectations to
the   search for neutral heavy resonances  in the $WW \rightarrow   e \nu \mu \nu$    decay channel  performed by the ATLAS Collaboration
using proton-proton  collision data at $\sqrt{s} = 13$ TeV and  corresponding to an integrated luminosity of 36.1  fb$^{-1}$~\cite{Aaboud:2018}.
To upgrade to the full Run 2 data set, we may use the recent results presented by the ATLAS Collaboration on the search of  heavy
resonances decaying into two vector bosons WW,  ZZ or WZ using collected data corresponding to  an integrated luminosity of
 139  fb$^{-1}$~\cite{Aad:2020}.
We will focus on the results for the search of  heavy neutral scalar resonance, called the Radion,   which appears in some
theoretical  models and which, indeed,  can decay into WW or ZZ  with a branching ratio approximatively given by  Eq.~(\ref{2.1}). 
 Moreover, the  Radion-like scalar resonaces couple to the Standard Model fermions  and gauge bosons with  strengths similarly to a heavy Higgs
boson. Considering that these heavy scalar resonances  have  a rather  narrow widths, we can consider the observed limits on the production processes as indicative of the production of a heavy Standard Model heavy Higgs boson in the narrow width approximation. \\ 
In  Fig.~\ref{Fig3} we display the observed limits at 95 \% confidence level on the heavy Higgs boson production cross section times the branching fraction  $Br(H \rightarrow WW + ZZ)$ for the gluon-gluon fusion  production mechanisms in
the narrow width approximation as reported in Ref.~\cite{Aad:2020}.  For comparison,  in Fig.~\ref{Fig3} we  also
report our estimate for the product of the gluon-gluon fusion production cross section  times the branching ratio for the decay of the heavy Higgs boson into two  vector bosons, after taking into account the reduced efficiency  as estimated in Eq.~({\ref{2.2}).
 Looking at  Fig.~\ref{Fig3}, it is evident  that, in the relevant mass range, our theoretical cross section is compatible
with  the observed  limits. On the other hand,  the observed gluon-gluon fusion cross section falls within the $\pm 2 \sigma$  ranges around the expected limit for the  Standard Model background-only hypothesis.
\section{The vector boson production mechanism}
\label{s-3}
\begin{figure}
\vspace{-0.0cm}
\includegraphics[width=0.5\textwidth,clip]{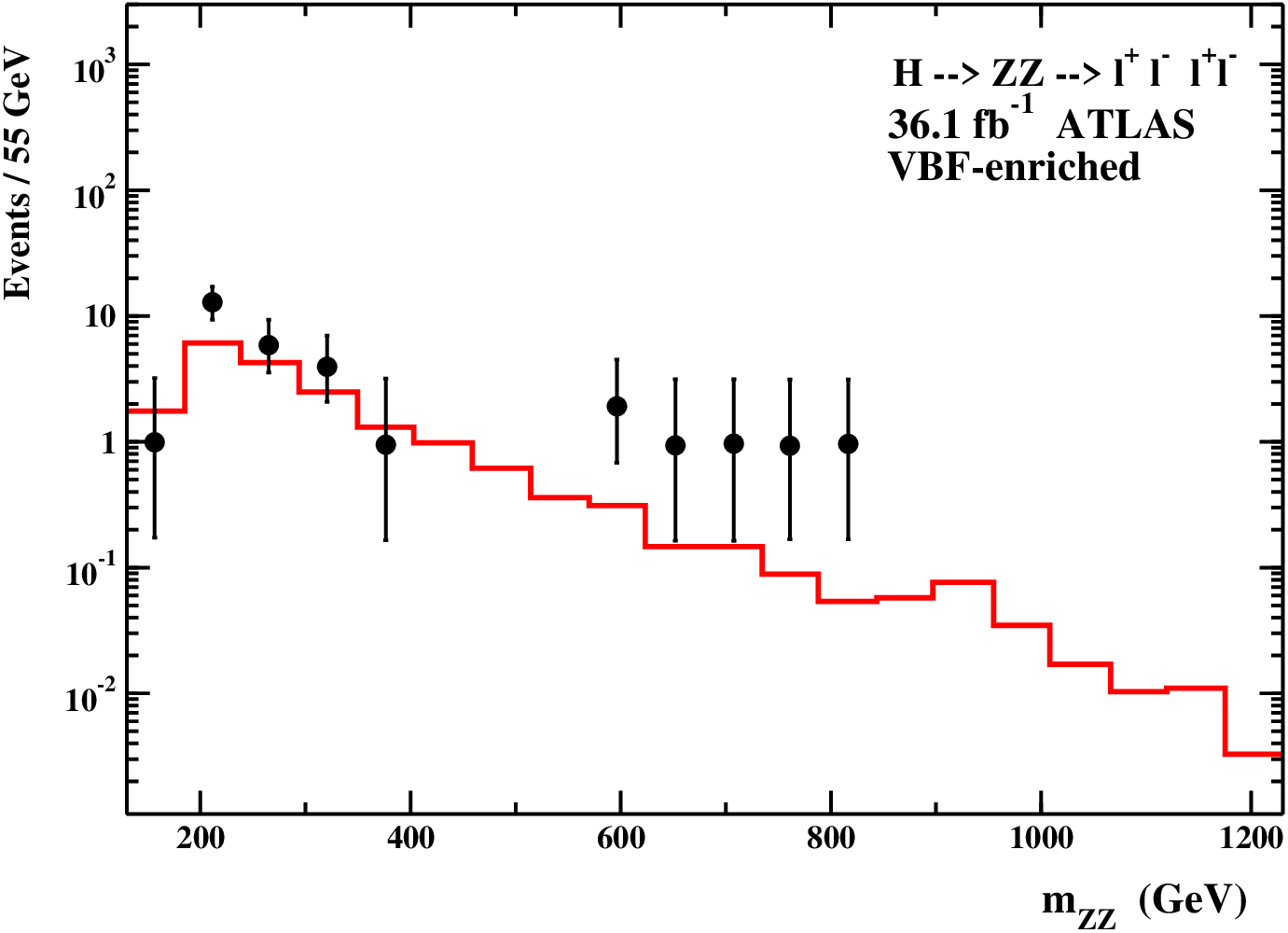}
\hspace{0.2 cm}
\includegraphics[width=0.5\textwidth,clip]{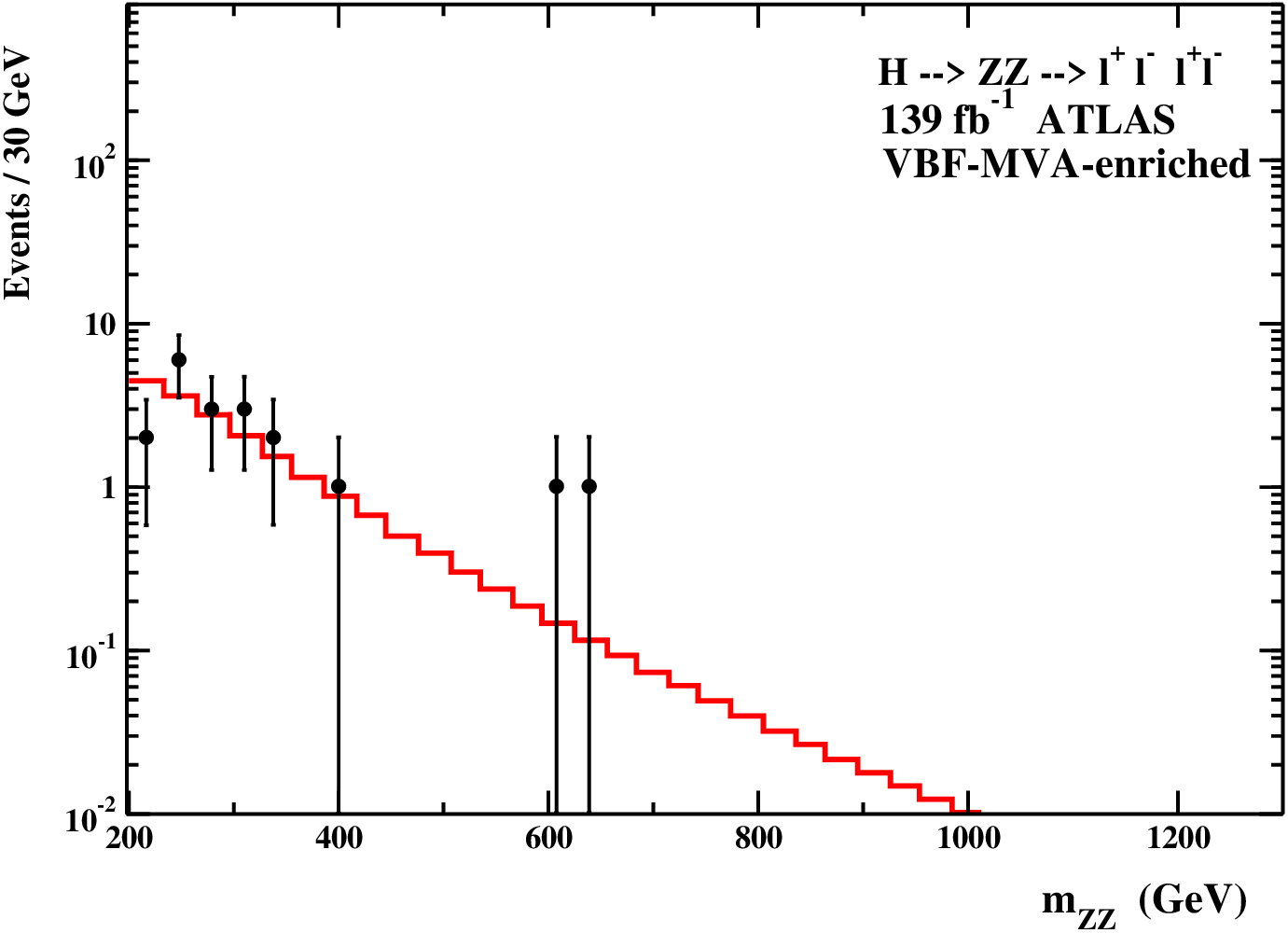}
\caption{\label{Fig4} 
Distribution of the invariant mass $m_{Z Z}$ for the VBF  processes $H \; \rightarrow ZZ \; \rightarrow \ell \ell \ell \ell$   ($\ell = e, \mu$)
corresponding  to  an  integrated luminosity of  ${\cal{L}} = 36.1 \,  fb^{-1}$ (left panel) and $139 \,  fb^{-1}$ (right panel).
The data  have been obtained  from Fig.~4 (b) in Ref.~\cite{Aaboud:2018} and  Fig.~2  (e) in Ref.~\cite{Aad:2021},  respectively.  
The (red) continuous lines  are the ZZ irreducible backgrounds.
}
\end{figure}
In the present Section we focus on the vector-boson fusion production mechanism. In   Fig.~\ref{Fig4} 
we display the invariant mass distribution for the golden channel for the VBF category corresponding to
36 fb$^{-1}$ (left panel) and 139 fb$^{-1}$ (right panel).  It is, now, worthwhile to comment on the effects of the
tightly  selection cuts applied by the ATLAS Collaboration  on the full Run 2 data. In fact, notwithstanding
the integrated luminosity has increased by a factor of about four, the number of events in the 
the high-mass region   $m_{Z Z}  > 600 \, $ GeV decreases from four to two. This corresponds to an
effective efficiency factor:
\begin{equation}
\label{3.1}
\eta_{eff}^{VBF} \; \simeq   \; 0.13 \; \; . 
\end{equation}
\begin{figure}
\vspace{-0.0cm}
\includegraphics[width=0.5\textwidth,clip]{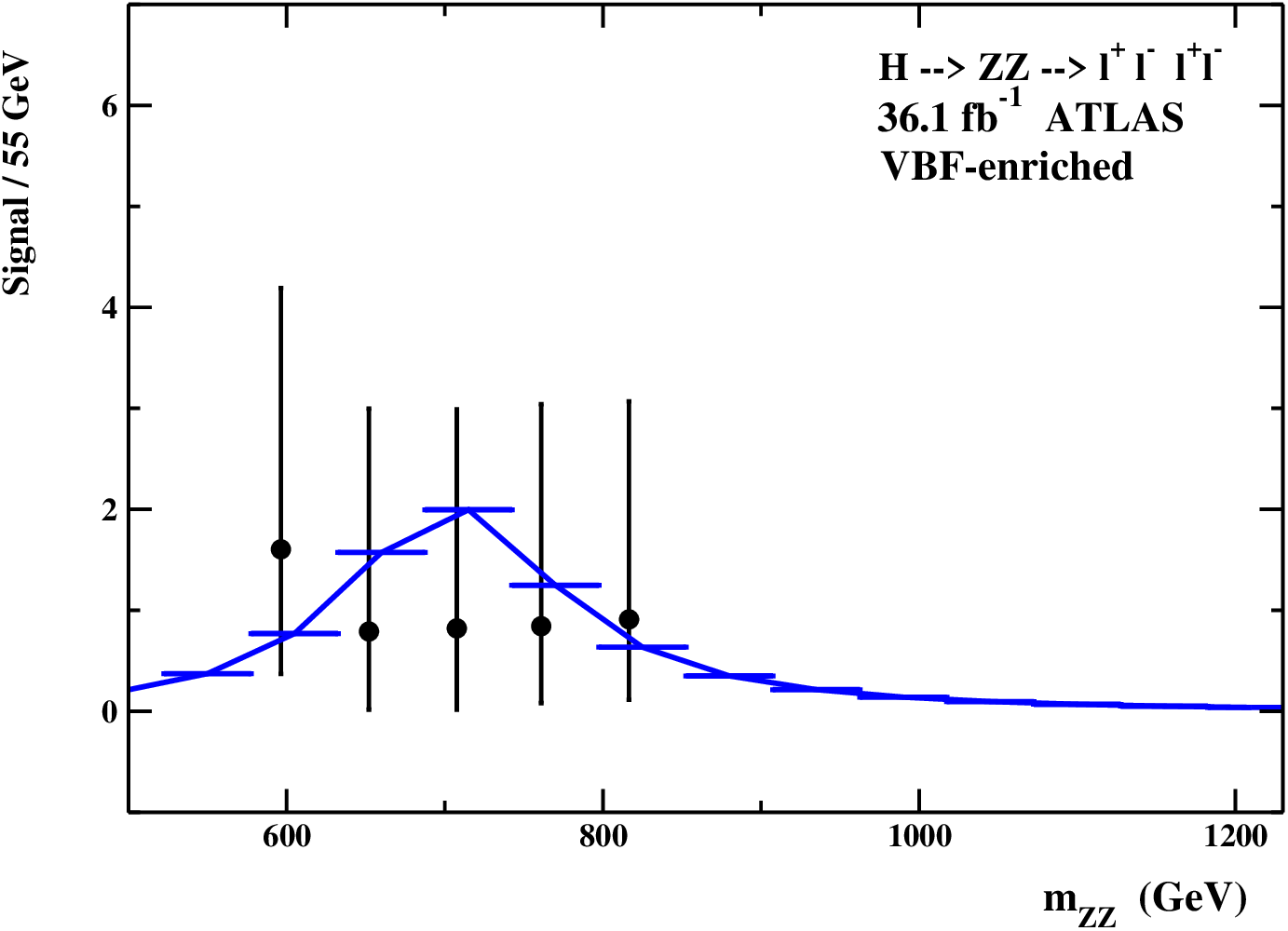}
\hspace{0.2 cm}
\includegraphics[width=0.5\textwidth,clip]{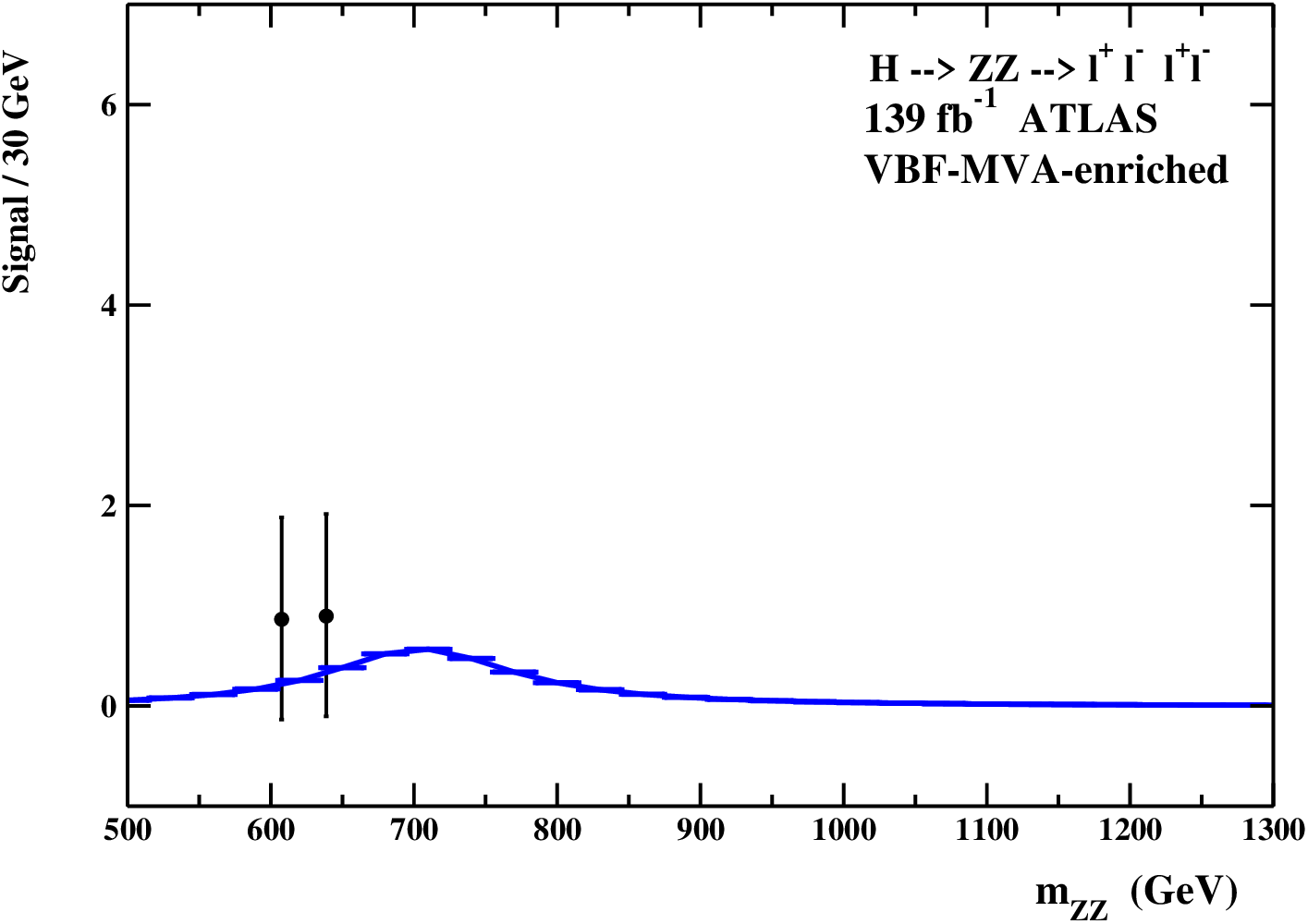}
\caption{\label{Fig5} 
Comparison to the Run 2  data of the  invariant-mass $m_{Z Z}$  distribution 
in the high-mass region   $m_{Z Z}  \gtrsim 600 \,  GeV$  for the  VBF processes
 $H \; \rightarrow ZZ \; \rightarrow \ell \ell \ell \ell$   ($\ell = e, \mu$) with integrated luminosity 36.1 fb$^{-1}$ (left panel)
 and 139 fb$^{-1}$ (right panel). The signal distribution  has been obtained from the relevant  ATLAS event distributions  by
subtracting the ZZ backgrounds.
 }
\end{figure}
In our opinion, these drastic effects derive from the event selection strategy that, for the full Run 2 data set,
is built with deep neural networks. It turns out that the neural networks are trained using the discriminating variables on simulated  events from
the decays of  a heavy Higgs bosons with  masses  $200$ GeV $ \leq$ $m_H \leq$   1400 GeV within the
narrow width approximation,  and the Standard Model ZZ background. We suspect that the event selection in the 
high invariant-mass region is  strongly biased due to the narrow width approximation and the strongly suppressed background of the Standard Model used in the training procedure of the neural networks. Indeed, the invariant-mass distribution from decays of our Higgs boson 
is a rather broad structure around 700 GeV that resemble more closely an almost continuous distribution than the  distribution arising
from the decays of a narrow-width Higgs boson.  Now, Fig.~\ref{Fig4}, right panel, shows that the Standard Model ZZ background is strongly suppressed  in the high invariant-mass region for Higgs production via  the vector-boson fusion process. Therefore, it is conceivable
that  eventual signals from our Higgs boson could have been rejected by the event selection  cuts. \\
\begin{figure}
\vspace{-0.3cm}
\begin{center}
\includegraphics[width=0.66\textwidth,clip]{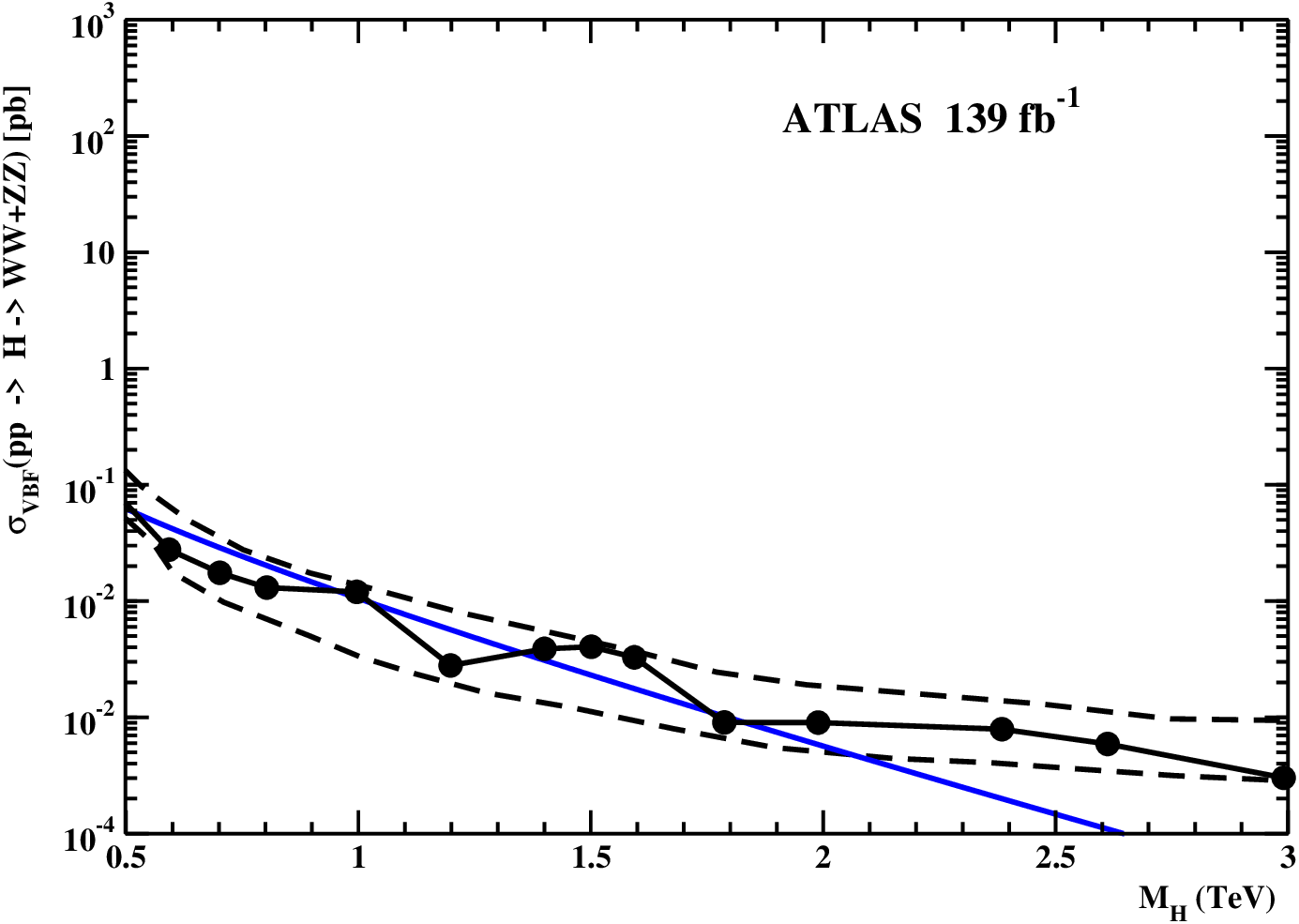}
\end{center}
\caption{\label{Fig6} 
Limits on the VBF production cross section times the branching fraction for the processes $pp \rightarrow H \rightarrow VV$, V = W,Z. 
The data have been taken from Fig.~12 (b) in Ref.~\cite{Aad:2020}. The dashed  lines demarcate  the 95 \% confidence level region
of the expected Standard Model background. Full circles represent the observed signal. The continuous (blu) line
is our theoretical estimate for the gluon-gluon fusion production cross section times the branching ratio 
Br(H $\rightarrow$ WW  + ZZ).
 }
\end{figure}
As in the previous Section,  from the event distributions we extract  the signal distributions by subtracting the   ZZ background
(see Fig.~\ref{Fig5}) and compare with the expected distribution. For the 2016 ATLAS data set,
we see that the expected signal histogram is perfectly compatible with the data  even though  the integrated luminosity is too low 
to support  the heavy Higgs boson with a significative statistical  significance.
As concern the full data set, in Fig.~\ref{Fig5}, right panel, we compare the expected signal distribution, scaled by the
effective efficiency factor Eq.~(\ref{3.1}),  to observations. Again, we see that there is not enough significancy to claim
an evidence.  This is confirmed more  concretely by comparing the expected and observed number of events in the high
invariant-mass region:
\begin{equation}
\label{3.2}
N^{exp}_{sign} ( m_{ZZ}  > 600 \,  {\text GeV}) \; \simeq  \; 1.76^{+1.36 }_{- 1.42}      \; , \; 
N^{th}_{sign} =  3.48   \; \;  \; \; \;   \;  VBF  \;  \;  {\cal{L}} = 139 \,  {\text fb}^{-1} \; \; \; .
\end{equation}
Finally, In  Fig.~\ref{Fig6} we display the observed limits at 95 \% confidence level on the heavy Higgs boson production cross section times the branching fraction  $Br(H \rightarrow WW + ZZ)$ for the vector-boson fusion  production mechanism in
the narrow width approximation as reported in Ref.~\cite{Aad:2020}.  For comparison,  we  also
report our estimate for the product of the VBF production cross section  times the branching ratio for the decay of the heavy Higgs boson into two  vector bosons, after taking into account the reduced efficiency  as estimated in Eq.~({\ref{3.1}).
 Looking at  Fig.~\ref{Fig6}, it is evident  that, in the relevant mass range, both our theoretical cross section and 
 the Standard Model background-only hypothesis are compatible with  the observed  limits.
\section{Summary and conclusions}
\label{s-4}
In our previous  paper we proposed to picture the Higgs condensate of the Standard Model as a quantum liquid analogous to
the superfluid helium. In this scenario   the Higgs condensate excitations behave as two Higgs bosons.
 The light Higgs boson was  already identified with the LHC narrow resonance at 125 GeV. As concern
the heavy Higgs boson, we  found preliminary evidence in our previous phenomenological analysis in the golden
channel of the preliminary LHC Run 2 data from ATLAS and CMS Collaborations.
In the present note we compared our proposal to the full Run 2 data set released  recently by the ATLAS Collaboration.
We do not found a clear statistical evidence for our Standard Model heavy Higgs. At beast we found a hint of a signal
in the gluon-gluon fusion Higgs production mechanism. We argued that  there is not enough sensitivity to detect the signal 
in vector-boson fusion mechanism mainly due to  tight event-selection cuts. In any case, we concluded that
 our theoretical proposal is still in accordance with the available  observations.

\end{document}